\documentclass[11pt,a4paper]{article}

\usepackage{graphicx}

\usepackage{cite}

\usepackage{hyperref}

\usepackage{geometry}
\title{Raychaudhuri Equation in an Anisotropic Universe with Anisotropic Sources}

\date{\today}

\author{Manabendra Sharma\footnote{email: manabendra@iiserb.ac.in}}

\begin{document}

\maketitle

\centerline{Department of Physics, IISER Bhopal, Bhopal - 462023, India}

\begin{abstract}
 In this paper we investigate the fate of the universe with an anisotropic background sourced by anisotropic matter. We see the behaviour of the Raychaudhuri Equation and investigate wheather a congrurence of time like geodesics focus to a point in a universe with Bianchi I background which is dictated by anisotropic sources like cosmic strings, domain walls, magnetic field and lorentz violating magnetic field each separately. Thus, Focussing theorem has been checked both for initially contracting and diverging universe with each of these anisotropic sources.
\end{abstract}

\section{Introduction}
The Raychaudhuri Equation Ref\cite{RaychaudhuriOriginal}, as we know,  is the starting point of the Singularity Theorem Ref\cite{LargeScaleStructure},Ref \cite{WaldBook}. The equation together with the Frobenius theorem gives us the Focussing theorem Ref\cite{Review} which predicts the fate of a congruence of geodesic provided the information of initial value (sign) of the expansion parameter be given. The evolution of the congruence depends on the energy conditions we choose. The focusing theorem, though not as sophisticated as the singularity theorem due to Hawking and Penrose, still gives us a crude idea of divergence or convergence of a bundle of geodesics given the initial conditions. The Raychaudhuri Equation as such is a geometrical identity. It is the Einstein's equation of gravity that connects geometry with the physics. Though the Raychaudhuri equation can accommodate any modified theory of grvity, in this piece of work, we will stick to GR limit. Recently, Ref \cite{RecentRCE} has reviewed the Focussing theorem for a Bianchi I background both for null and time like congruence in the GR limit and also have extended their work to some $f(R)$ model.

In this paper we investigate the focusing theorem for anisotropic sources in an anisotropic background. We restrict our discussion to a time like congruence only. We choose a Bianchi I metric with a residual planar symmetry and correspondingly the anisotropic sources like cosmic string, domain walls, magnetic fields and Lorentz violating magnetic field . This is well motivated from the work of Ref\cite{Alurietal}. This small piece of work is divided into two sections. In section \ref{sec:RCEbianchiI}, we introduce the Einstein field equation and the Raychaudhuri Equation. This is followed by section \ref{sec:ftheorem}, wherein we investigate the details of the Focussing theorem for a congruence of time like geodesic in an anisotropic, homogenous background with anisotropic sources cosmic string (CS), domain walls (DW), Lorentz violating magnetic field (LVMF) and magnetic field (MF).

\section{Raychaudhuri Equation for a Bianchi I metric with anisotropic sources}\label{sec:RCEbianchiI}
We consider a Bianchi I metric as a background with residual symmetry which takes the form

\begin{equation}
ds^2 = dt^2  - a(t)^2 dx^2 - b(t)^2 (dy^2 + dz^2).
\label{eq:metric}
\end{equation}

The Einstein field equation for this system looks like

\begin{eqnarray}
\frac{\mathrm{d} H}{\mathrm{d} t}&=&-H^2-\frac{2}{3}h^2-\frac{1}{6}(\rho+2p_b+p_a) \,, \\\label{eq:rce}
\frac{\mathrm{d} h}{\mathrm{d} t}&=&-3hH+\frac{1}{\sqrt3}(p_b-p_a) \,, \\ 
\label{eq:efet}
\end{eqnarray}

with,

\begin{equation}
H^2=\frac{\rho}{3}+\frac{h^2}{3}. 
\label{eq:modifiedfd}
\end{equation}

and

\begin{equation}
\frac{\mathrm{d} \rho}{\mathrm{d} t}=-3H\left ( \rho + \frac{p_a+2p_b}{3} \right )-\frac{2h}{\sqrt 3}\left ( p_b-p_a \right ) 
\label{eq:ContinuityEquation}
\end{equation}

where, $H=\frac{H_a+2H_b}{3}$ and $h=\frac{H_b-H_a}{\sqrt 3}$ are respectively the average Hubble parameter and the gravitional shear. The energy momentum tensor has a structure of the form $T_{\mu}^{\nu}=\left ( \rho, -p_a,-p_b,-p_b \right )$ and the equation of state for each of the matter considered are listed on the Table~\ref{tabl:Matter}.

Here Eq.~[\ref{eq:modifiedfd}] is the modified Friedmann equation and Eq.~[\ref{eq:ContinuityEquation}] is the continuity equation. 
The equation Eq.~[\ref{eq:rce}], is basically the Raychaudhuri equation in the absence of the rotating term for a bundle of a time like geodesic in the GR limit. Here, the last term consisting of $\rho+p_a+2p_b$ is the modified active gravitational mass density \cite{Ellis} for a Bianchi I background with anisotropic sources. Defining an averaged equation of state parameter $w=\frac{w_a+2w_b}{3}$ the active gravitional mass density can be phrased as $\rho+3w\rho$.

Now, we express our above set of equation in a frame where $t'=\frac{H}{H_0}$, $H'=\frac{H}{H_0}$ $h'=\frac{h}{h_0}$, $\rho'=\frac{\rho}{\rho_0}$ and $\Omega_0=\frac{\rho_0}{3H^2}$ and $\epsilon=\frac{h_0}{H_0}$. Thus defined, the new set of scaled equation looks like the following:

\begin{eqnarray}
\dot{H}'=-{H'}^2-\frac{2}{3}\epsilon ^2h'^2-\frac{1}{2}\left ( 1+w_a+2w_b \right )\Omega _0\rho ' \,, \nonumber\\
\dot{h'}=-3h'H'+\sqrt 3 \left ( w_b-w_a \right )\epsilon \Omega _0 \rho ' \,, \nonumber\\
\dot{\rho '} = -3\left ( 1 + \frac{w_a+2w_b}{3}\right )H'\rho '-\frac{2\left ( w_b-w_a \right )}{\sqrt 3}\epsilon h'\rho ' \,, \nonumber
\end{eqnarray}

\begin{table}
 \centering
 \begin{tabular}{|l|l|l|l|l|r|}
  \hline
   Matter                &  $w_a$    &  $w_b$   &    $w=\frac{w_a+2wb}{3}$        &  $1+3w$       \\
  \hline 
   Cosmic String         &    -1     &   0     &       $\frac{-1}{3}$             &   0            \\
   Domain Walls          &     0     &  -1     &       $\frac{-2}{3}$             &   -1            \\
   LVMF                  &     1     &   0     &       $\frac{1}{3}$              &   2             \\
   MF                    &    -1     &   1     &       $\frac{1}{3}$              &   2             \\
   
  \hline
 \end{tabular}
 \caption{Equation of state parameter}
 \label{tabl:Matter}
\end{table}

\begin{figure}
 \centering
 $
 \begin{array}{c c}
  \includegraphics[width=0.54\textwidth]{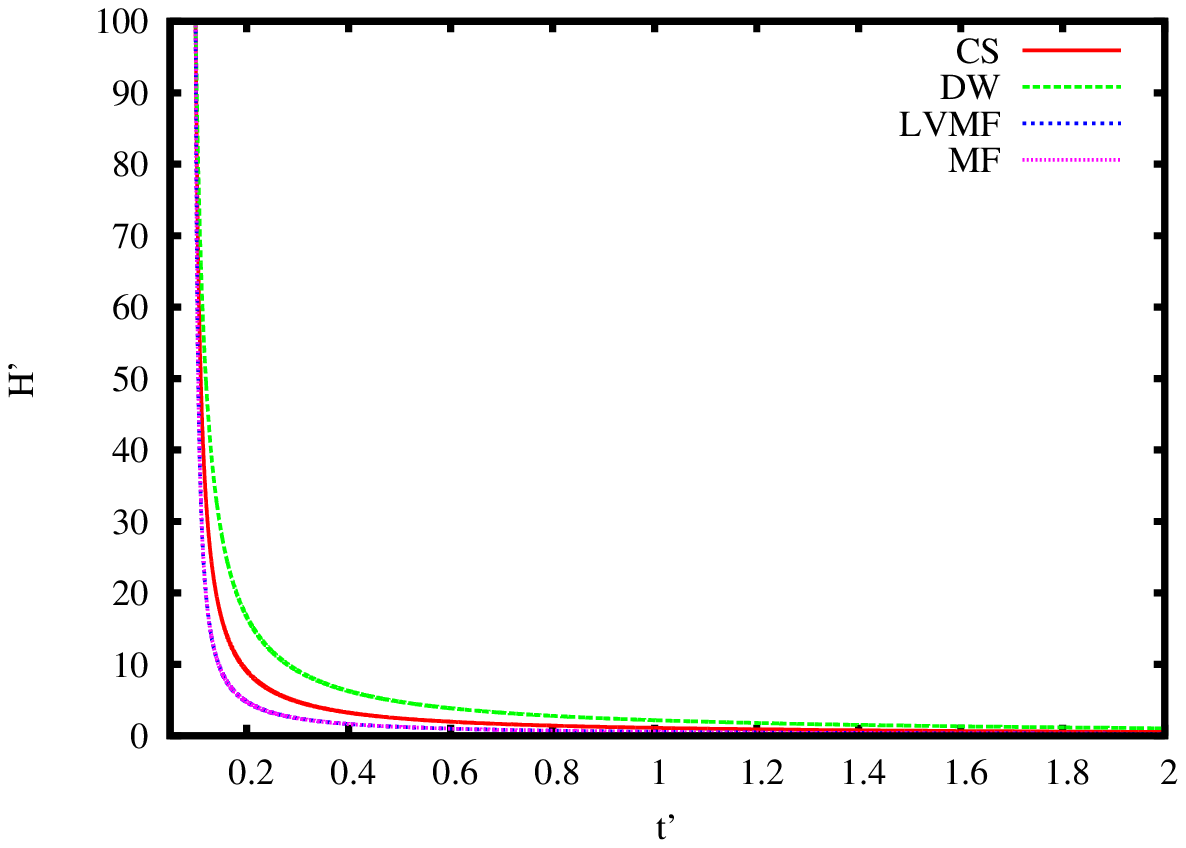} &
  \includegraphics[width=0.54\textwidth]{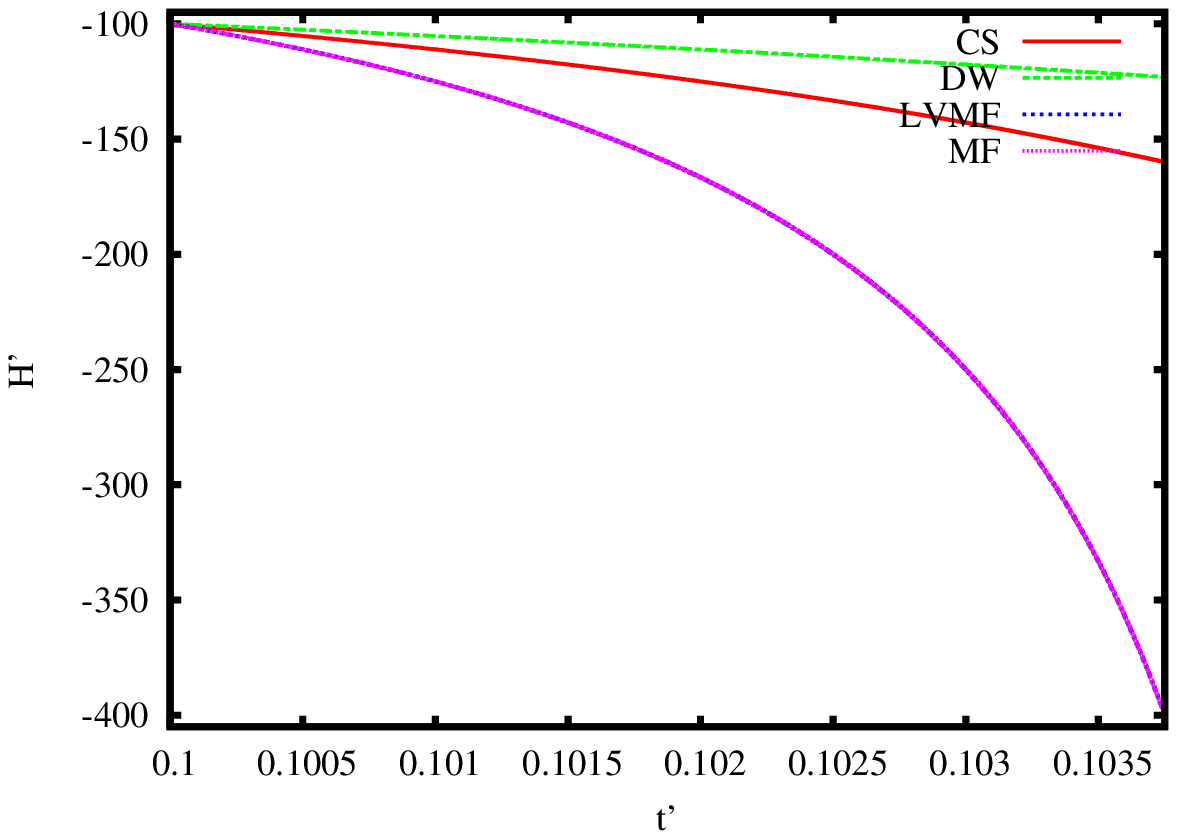}
 \end{array}        
 $
 \caption{\emph{Left} Plot for Exp. Universe, $H'$ vs. $t'$ and \emph{Right} plot for Cont. Universe $H'$ vs. $t'$ }
 \label{fig:Hubble}
\end{figure}

\begin{figure}
 \centering
 $
 \begin{array}{c c}
  \includegraphics[width=0.54\textwidth]{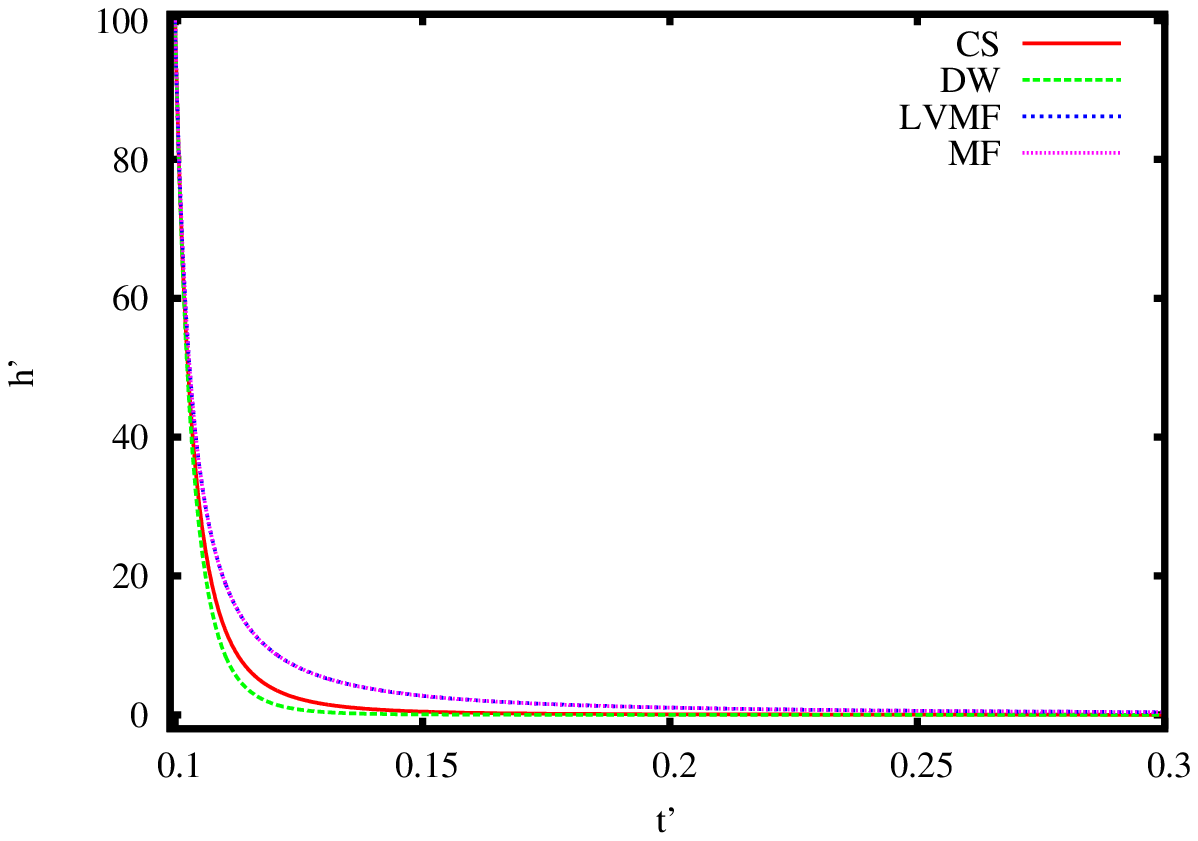} &
  \includegraphics[width=0.54\textwidth]{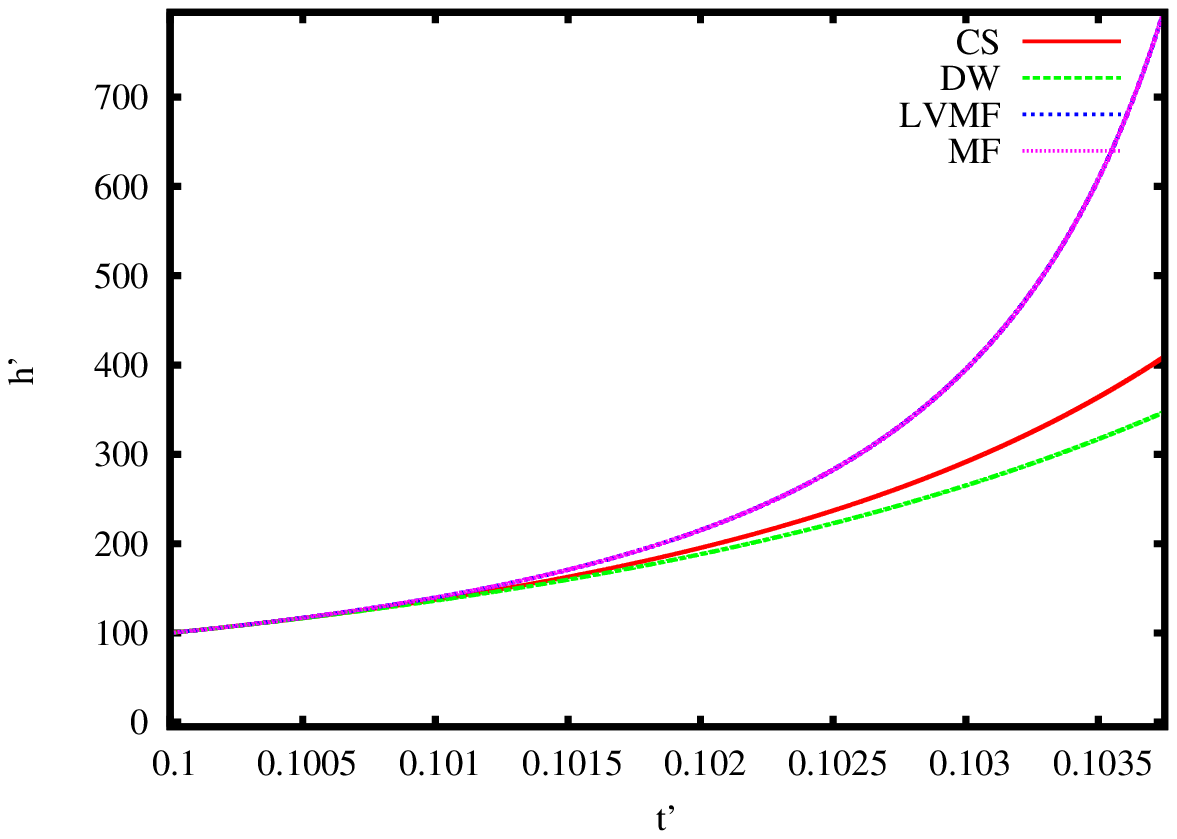}
 \end{array}        
 $
 \caption{\emph{Left} for Exp. Universe, $h'$ vs. $t'$ and \emph{Right} for Cont. Universe, $h'$ vs. $t'$ }
 \label{fig:Shear}
\end{figure}

\begin{figure}
 \centering
 $
 \begin{array}{c c}
  \includegraphics[width=0.54\textwidth]{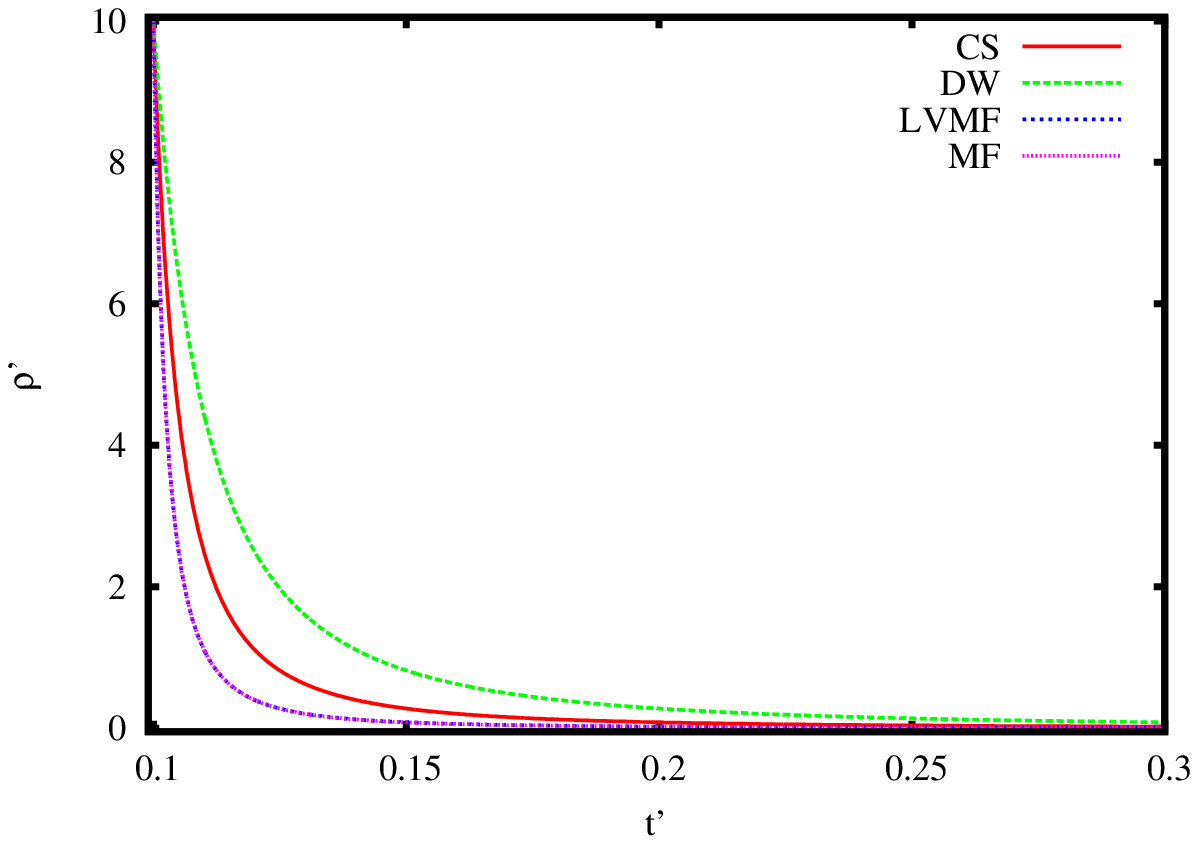} &
  \includegraphics[width=0.54\textwidth]{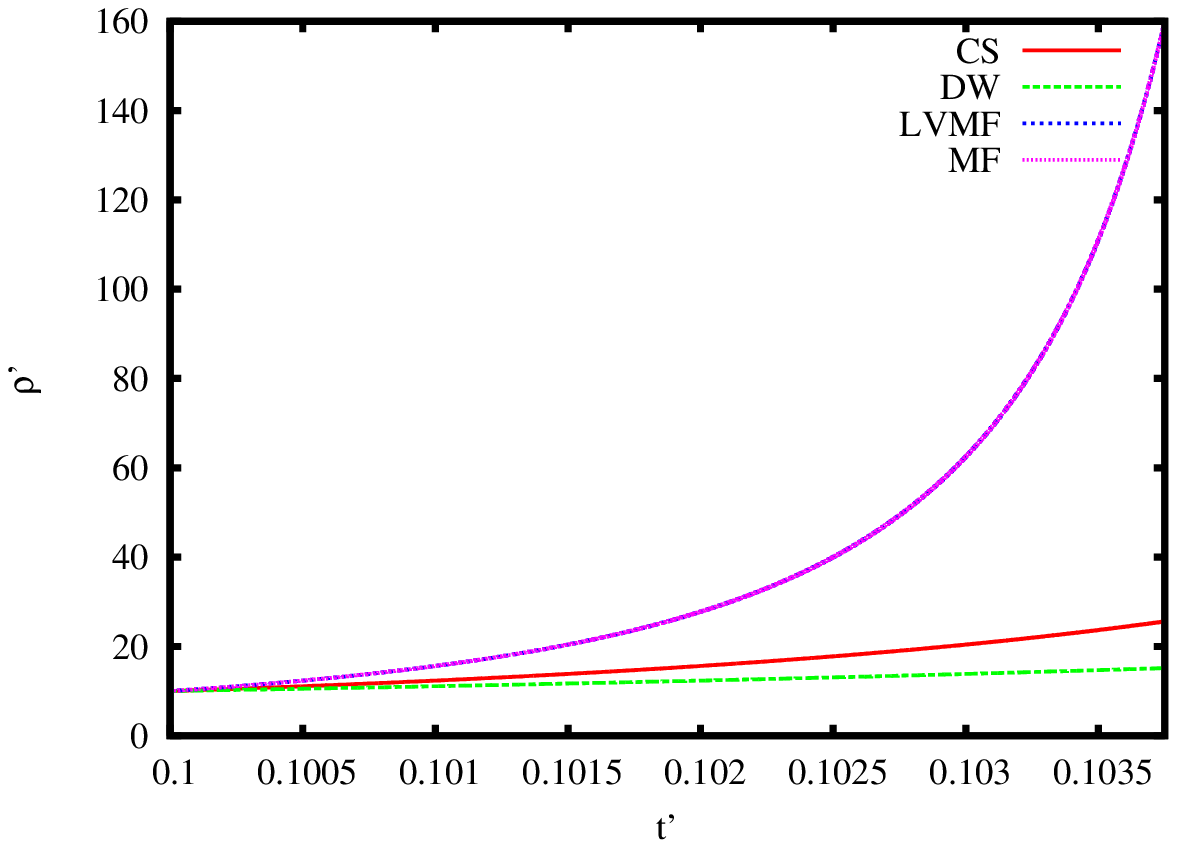}
 \end{array}        
 $
 \caption{\emph{Left} for Exp. Universe, $\rho'$ vs. $t'$ and \emph{Right} for Cont. Universe, $\rho'$ vs. $t'$ }
 \label{fig:Energy Density}
\end{figure}

\begin{figure}
 \centering
 $
 \begin{array}{c c}
  \includegraphics[width=0.54\textwidth]{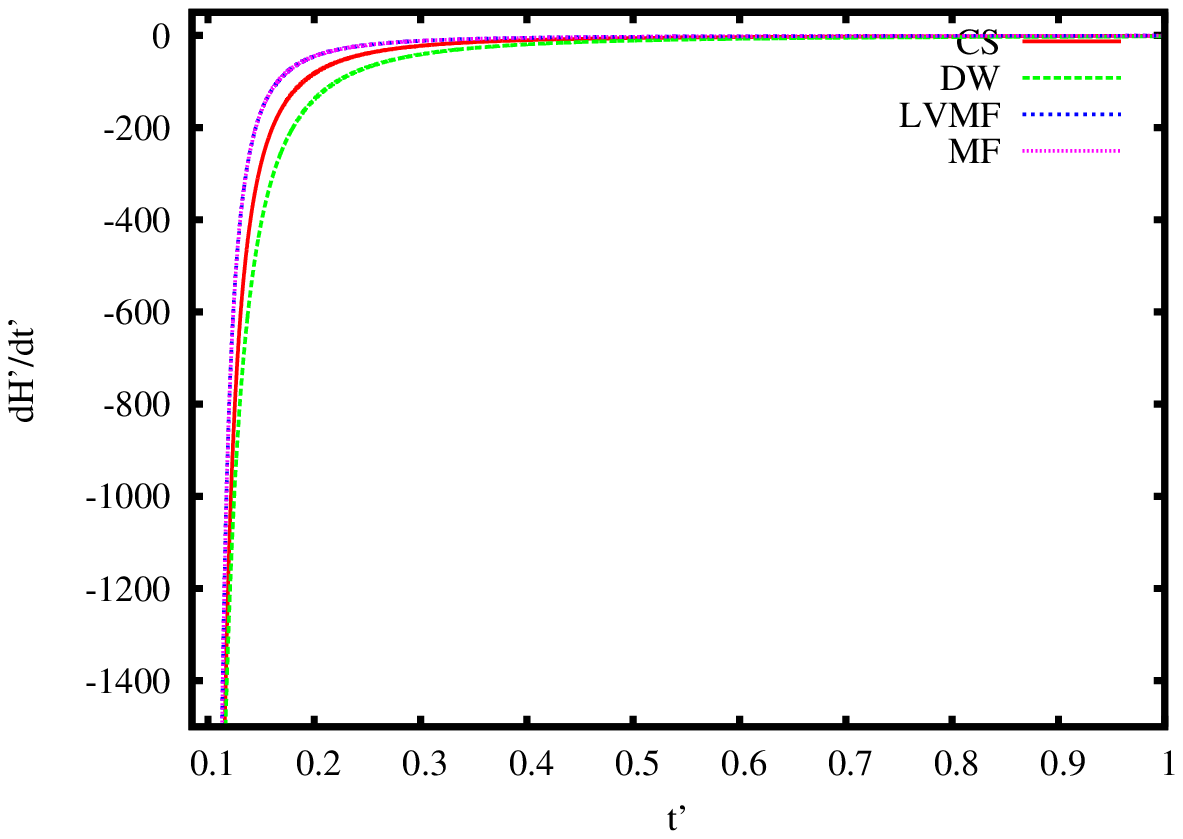} &
  \includegraphics[width=0.54\textwidth]{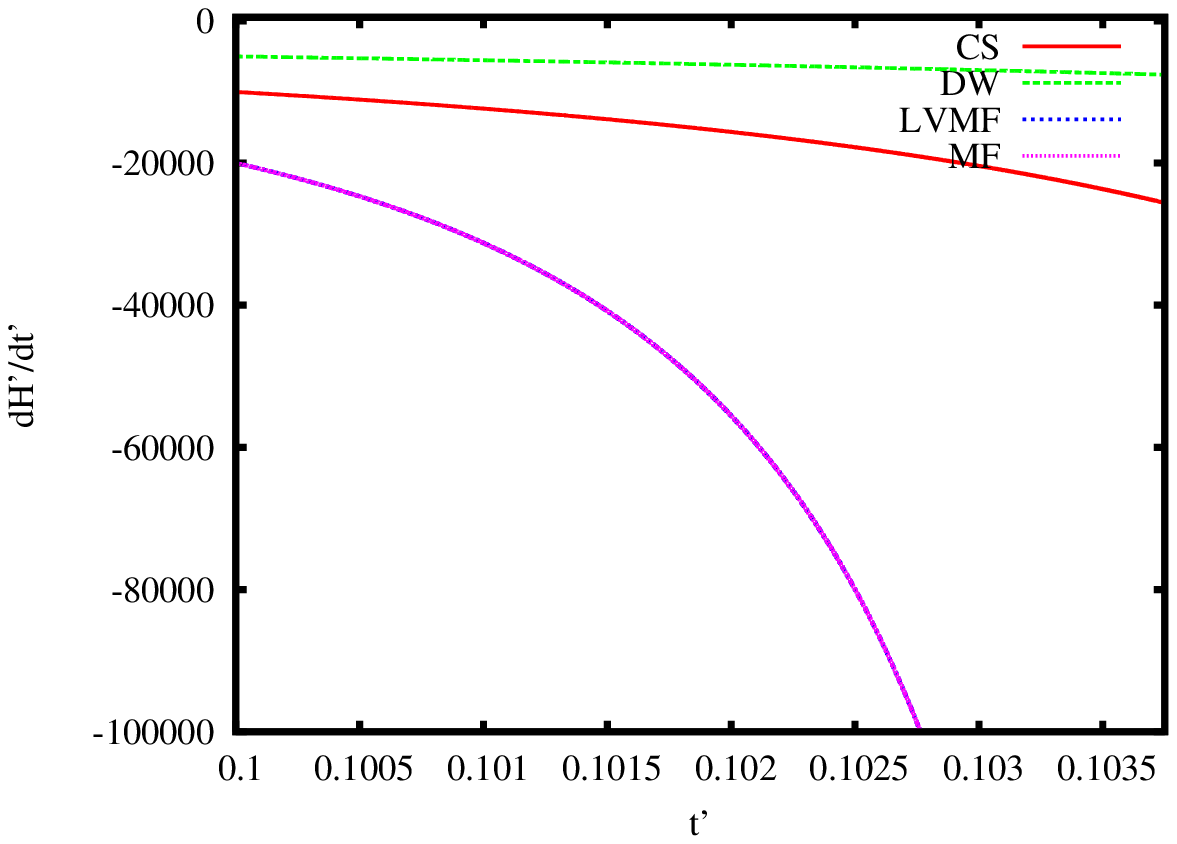}
 \end{array}        
 $
 \caption{\emph{Left} plot for Exp. Universe, $\frac{\mathrm{d} H'}{\mathrm{d} t'}$ vs. $t'$ and \emph{Right} plot for Cont. Universe, $\frac{\mathrm{d} H'}{\mathrm{d} t'}$ vs. $t'$ }
 \label{fig:Rate of Hubble}
\end{figure}

\begin{figure}
 \centering
 $
 \begin{array}{c c}
  \includegraphics[width=0.54\textwidth]{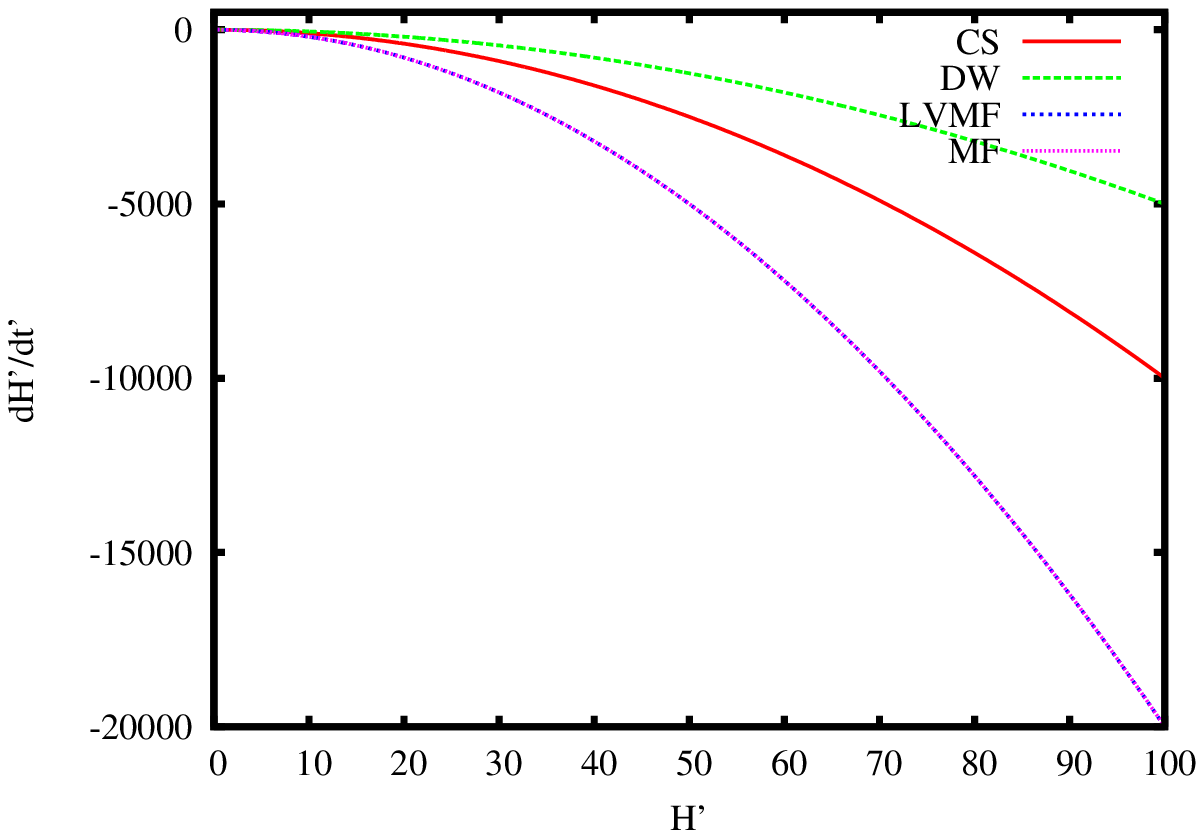} &
  \includegraphics[width=0.54\textwidth]{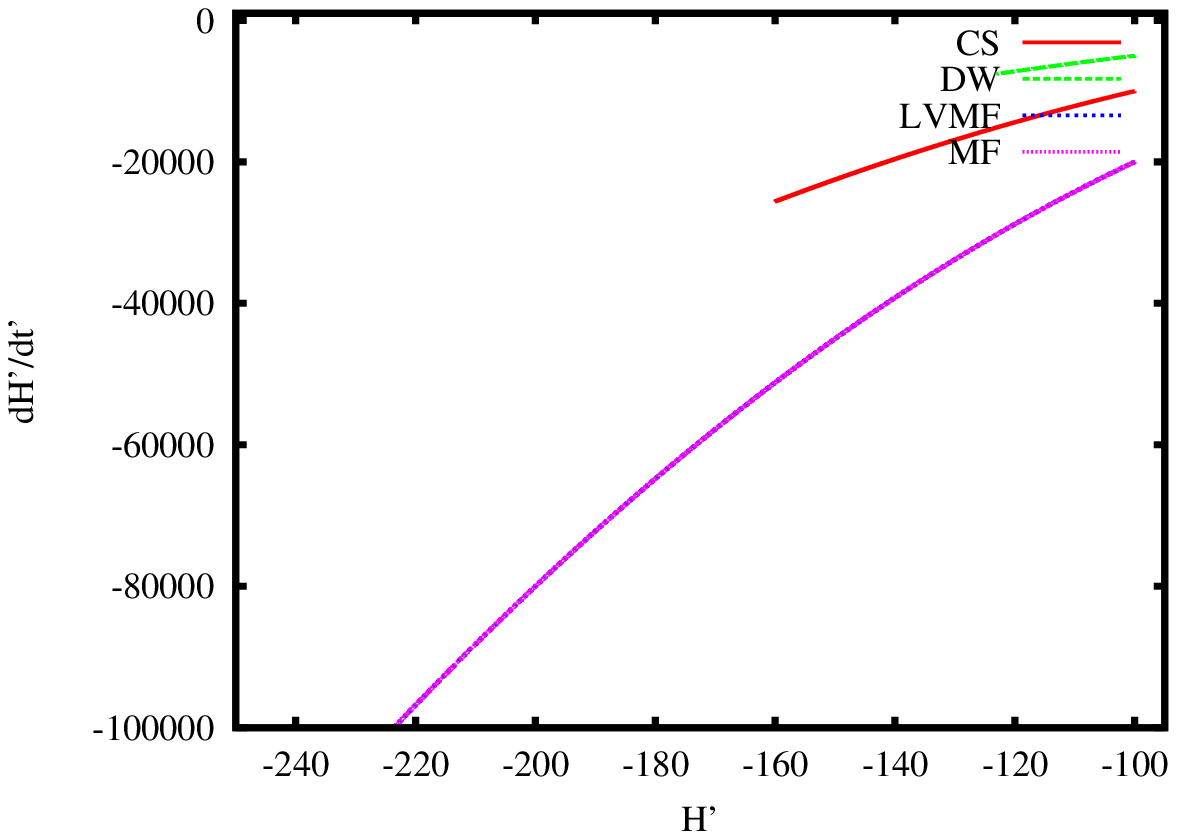}
 \end{array}        
 $
 \caption{\emph{Left} plot for Exp. Universe, $\frac{\mathrm{d} H'}{\mathrm{d} t'}$ vs. $H'$ and \emph{Right} plot for Cont. Universe, $\frac{\mathrm{d} H'}{\mathrm{d} t'}$ vs. $H'$ }
 \label{fig:Hubble phase space}
\end{figure}

\newpage
\section{Focusing Theorem and discussion of results}\label{sec:ftheorem}

The Focussing theorem says that an initially diverging congruence's would diverge less rapidly in the future, while an initially converging congruence will converge more rapidly in the future provided certain energy condition holds good. Thus, in other words, it says gravitation is an attractive force when strong energy condition is obeyed and consequently geodesics get focused.

The Hubble parameter signifies the fractional rate of change of volume:

\begin{equation}
H=\frac{\Delta V}{V\Delta t}
\end{equation}
 hence, a positive value of $H$ would imply an expanding universe and thus, a diverging congruence whereas a negative value of the same gives a contracting one leading to convergence of geodesics.

In this section, we look at the evolution of the Hubble parameter along with other field variables. The set of scaled EFE have been evolved in t' frame, numerically, for both negative and positive initial values oh H. All the plots shown here are for the same set of parameters $h_{initial}=100.0$, $\epsilon=0.00001$, $\Omega_0=1000$ except $H_{initial}=+100.0$ for expansion and $H_{initial}=-100.0$ for contraction. The values of $\rho$ has been calculated using the modified Friedmann.

With these, Fig.~[\ref{fig:Hubble}] shows the evoltion of the Hubble parameter for expansin on the left while contraction on the right for each of the anisotropic sources considered. Whereas the Hubble parameter, which is proportional to fractional rate of change of volume, goes assymptotically to a constant value of zero Ref\cite{Alurietal} for expansion, on the other hand, it goes to negative infinity for a contracting universe. This means, though, future singularity is not achieved in the case of expansion but it can not be avoided for an initially contracting one. Thus, a congruence of time like geodesics converge to a point in future as predicted by the Focussing theorem. Fig.~[\ref{fig:Rate of Hubble}] and Fig.~[\ref{fig:Hubble phase space}] depicts the same showing slope of the Hubble parameter evolving to zero in the case of expansion and steeply cascading down to negative infinity for each component of matter considered which in turn confirms focussing.
The relative slowness or fastness of decrease of the Hubble parameter for each component of the anisotropic matter depends largely on the equation of state parameter.
A larger positive value of averaged equation of state parameter $w$ would fasten the evolution of $H$ whereas a negative value of the same would slow down the rate of H, which can easily be seen from the plots and from the Table~\ref{tabl:Matter}. The fact that the plot of magnetic feild and Lorentz violating magnetic field  overlaps is because of the very reason that they bear the same value of averaged equation of state parameter $w$ despite having different individual values of $w_a$ and $w_b$. This clearly signifies the importance of calling the $(\rho+w_pa+2p_b)$ to be the active gravitational mass density.

The evolution of gravitational shear h, which is nothing but the difference of Hubble parameter in two different direction are shown in Fig.~[\ref{fig:Shear}]. The shear parameter for an expanding universe shows a "normal" behaviour. That is, the universe isotropizes asymptotically for each component of the anisotropic matter. The relative behaviour of h for each component of the anisotropic matter and a detailed extensive study of fixed point analysis have been carried out in Ref\cite{Alurietal}. The interesting  point to look at in these paper is for an initially contracting universe wherin a focussing of congruence in future can not be avoided. The behaviour of h in such case has been depicted on the right of Fig.~[\ref{fig:Shear}] which shows an anisotropic collapse of the universe. Thus, as the initially contracting universe progresses towards a future focusing the difference of Hubble parameter in two different direction blows up. In addition to this, the relative slowness or fastness of the rise of shear for each component of matter depends largely on the difference of pressure, parametrized as $w_b-w_a$, in two different directions.

Now, as we show the decrease of energy density with the expansion of the universe, Fig.~[\ref{fig:Energy Density}], indicates a sharp rise of the energy density $\rho$ in the contracting one for each component of anisotropic matter. This is expected because energy density is inversely proportional to the volume. Thus, a forever contracting universe becomes infinetly densed as it approaches to the focussing point.

\section{conclusion}
The focussing theorem has been studied for a Bianchi I background dictated by anisotropic sources such as cosmic string, domain walls, Lorentz violating magnetic field and magnetic field. That is, the fate of the congruence of a time like geodesic have been studied both for an eapanding and a contracting universe and it has been found that the congruence focus to a point for initially contracting univervse as there is no adequate matter to bounce back. Thus, for a contracting universe as governed by the anisotropic sources considered it meets an anisotropic collapse as the shear and energy density blows up to a singularity.

\end{document}